\RequirePackage{etoolbox}
\newtoggle{daconf}

\iftoggle{daconf}{%
  \documentclass[12pt,conference,compsoc,onecolumn]{drivingassessment}
  \usepackage[round,sort,authoryear]{natbib}
  \bibliographystyle{abbrvnat}
  \setcitestyle{authoryear}
  \setlength{\bibsep}{6pt plus 0.3ex}
}{%
  \documentclass[conference]{IEEEtran}
  \usepackage[square,sort,numbers]{natbib}
  \bibliographystyle{abbrvnat}
}

\RequirePackage{snapshot}

\usepackage{caption}
\usepackage{subcaption}

\usepackage{moreverb}
\usepackage{verbatim}

\usepackage{color}
\usepackage{balance}

\usepackage{parskip}
\usepackage[inline]{enumitem}
\usepackage{latexsym}
\usepackage[pdftex]{graphicx}

\usepackage[sharp]{easylist}

\usepackage{multirow}

\usepackage{inconsolata}
\newcommand{\weblink}[1]{\path{#1}}

\usepackage[usenames,dvipsnames,table]{xcolor}

\usepackage{amsmath}
\usepackage{amsfonts}
\usepackage{amssymb}
\usepackage{amsthm}

\usepackage{algorithmic}
\usepackage{algorithm}
\usepackage{cases}

\renewcommand{\eqref}[1]{(\ref{eq:#1})}

\newcommand{\figref}[1]{Fig.~\ref{fig:#1}}

\usepackage{fancyhdr}
\fancypagestyle{firststyle}
{
  \fancyhf{}
  
   \fancyfoot[L]{
     \IEEEauthorrefmark{4} Corresponding author: \texttt{fridman@mit.edu}
   }
}

\begin{document}

\title{Eye Contact Between Pedestrians and Drivers}
  

\newcommand{\authorspace}{\hspace{0.2in}}
\author{
  Dina AlAdawy\IEEEauthorrefmark{1} \authorspace
  Michael Glazer\IEEEauthorrefmark{1} \authorspace
  Jack Terwilliger\IEEEauthorrefmark{1} \authorspace
  Henri Schmidt\IEEEauthorrefmark{1}\vspace{0.03in}
  \\
  Josh Domeyer\IEEEauthorrefmark{2} \authorspace
  Bruce Mehler\IEEEauthorrefmark{1} \authorspace
  Bryan Reimer\IEEEauthorrefmark{1} \authorspace
  Lex Fridman\IEEEauthorrefmark{1}\IEEEauthorrefmark{4}\vspace{0.15in}
  \\
  \IEEEauthorblockA{\IEEEauthorrefmark{1}Center for Transportation and Logistics,\\Massachusetts Institute of Technology}\\
  \IEEEauthorblockA{\IEEEauthorrefmark{2}Toyota Collaborative Safety Research Center}
}


\maketitle

\begin{abstract}
  When asked, a majority of people believe that, as pedestrians, they make eye contact with the driver of an approaching vehicle when making their crossing decisions. This work presents evidence that this widely held belief is false. We do so by showing that, in majority of cases where conflict is possible, pedestrians begin crossing long before they are able to see the driver through the windshield. In other words, we are able to circumvent the very difficult question of whether pedestrians choose to make eye contact with drivers, by showing that whether they think they do or not, they can't.  Specifically, we show that over 90\% of people in representative lighting conditions cannot determine the gaze of the driver at 15m and see the driver at all at 30m. This means that, for example, that given the common city speed limit of 25mph, more than 99\% of pedestrians would have begun crossing before being able to see either the driver or the driver's gaze. In other words, from the perspective of the pedestrian, in most situations involving an approaching vehicle, the crossing decision is made by the pedestrian solely based on the kinematics of the vehicle without needing to determine that eye contact was made by explicitly detecting the eyes of the driver.
  %
\end{abstract}


\thispagestyle{firststyle}
\setlength{\footskip}{20pt}

\section{Introduction}\label{sec:introduction}

When asked, most people will say that they make eye contact with the driver of an approaching vehicle before crossing
the road (see Survey section). This work provides evidence that this belief is largely false, and consequently, that
people misattribute the source of their crossing decisions.  We focus on the subclass of pedestrian-vehicle interactions
where collision is possible as a consequence of non-verbal miscommunication; specifically, situations where a vehicle is
decelerating for a pedestrian and there is no crosswalk or signal. While in most road contexts, the law clearly defines
who has the right of way, the reality of social-interaction in these contexts is that laws are often bent and
broken. Just like many drivers partake in regularly exceeding the speed limit, many pedestrians partake in regularly
crossing the street when and where they are not legally allowed to.

In this world, where formal rules of the road are not strictly followed, companies and researchers are beginning to test
autonomous vehicles on public roads. The important question becomes: What are the cues that pedestrians and vehicles use
in ultimately resolving these situations so that we can design autonomous vehicle control algorithms that interact
successfully with pedestrians. For the pedestrian, despite the aforementioned common misconception, the crossing
decision appears to be primarily based on vehicle kinematics \citep{brewer2006exploration}. In fact, one explanation for
the findings in this work is that pedestrians form models of driver intent by observing vehicle kinematics, and because
of forming such a model believe that they in-fact ``see'' the driver. We conduct a survey and two experiments that
provide evidence that they do not see the driver.

Our work has several actionable takeaways for the design of safe human-centered autonomous vehicle control algorithms
\citep{fridman2018humancentered}. First, the general inability to see the presence of a driver in an approaching
vehicle means that an autonomous vehicle that has no driver can still, in large part, participate in the intricate
pedestrian-vehicle interaction without the pedestrian being directly aware of the driver's absence. Second, this work implies
that vehicle kinematics is in fact the primary form of non-verbal cuing from vehicle to pedestrian. Acceleration and
deceleration of the vehicle is something that an autonomous vehicle can control precisely and directly learn from real-world interactions.


\newcommand{\lightvspace}{\vspace{0.05in}}
\newcommand{\lightwidth}{0.325\textwidth}
\newcommand{\lightsub}[1]{%
\begin{subfigure}{\lightwidth}%
\includegraphics[width=\linewidth]{images/car_small_select/#1.jpg}%
\end{subfigure}%
}

\begin{figure*}[!ht]
  \centering  
  \lightsub{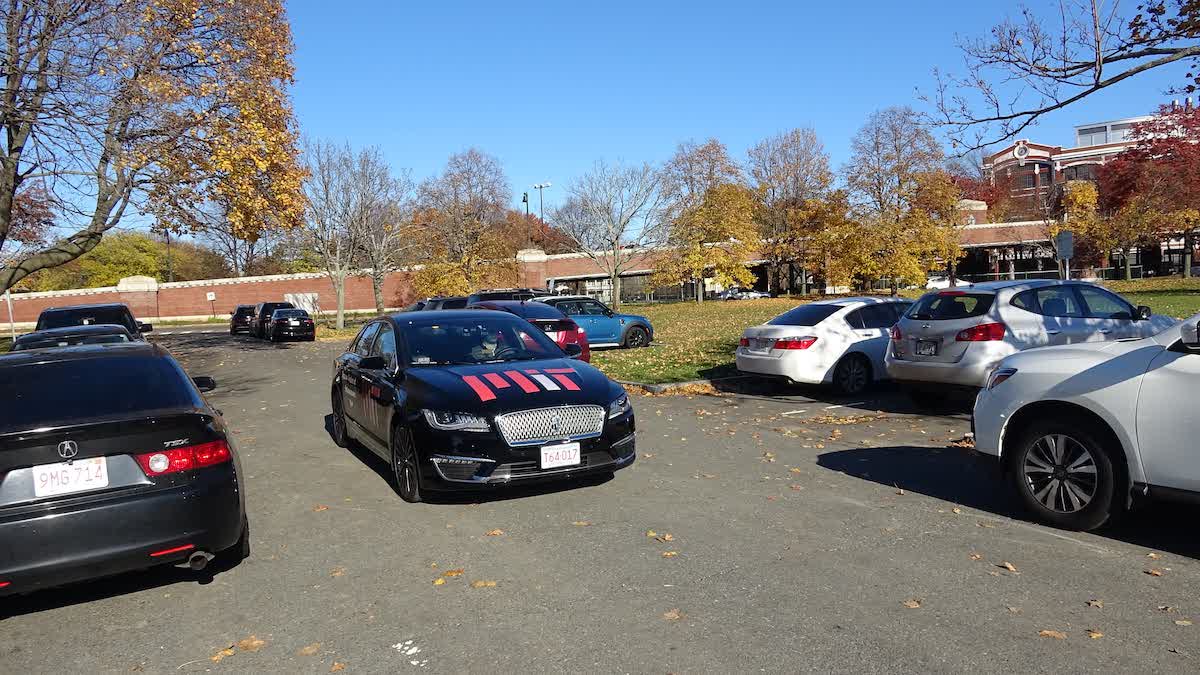}
  \lightsub{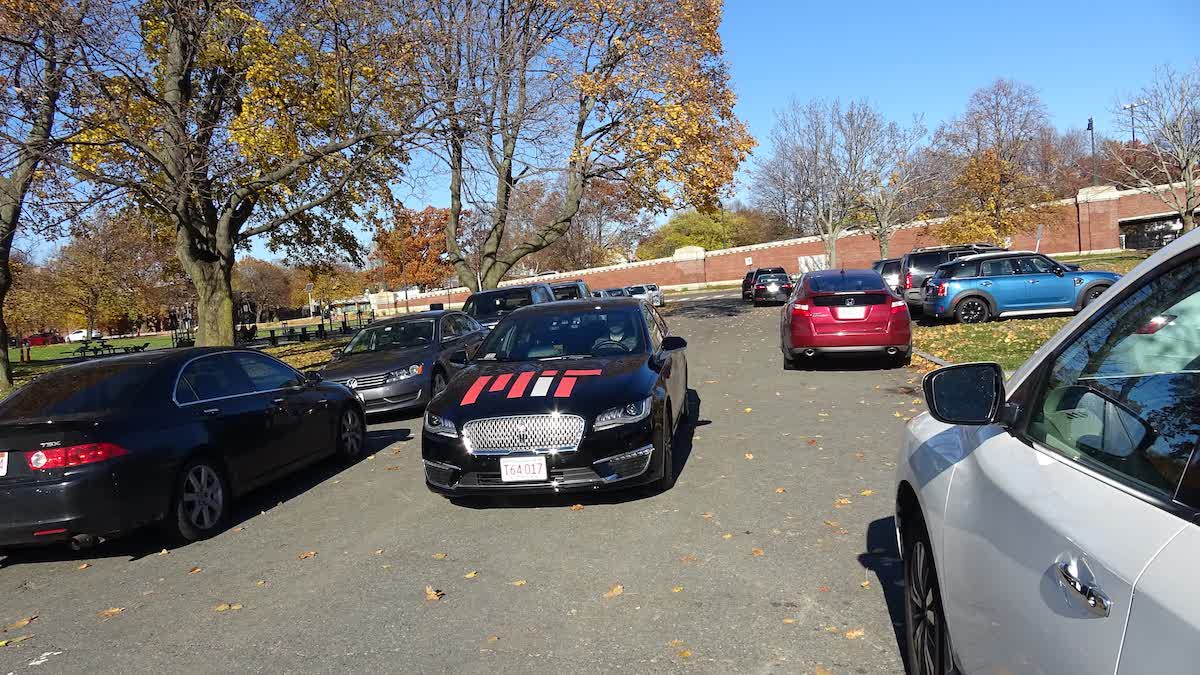}
  \lightsub{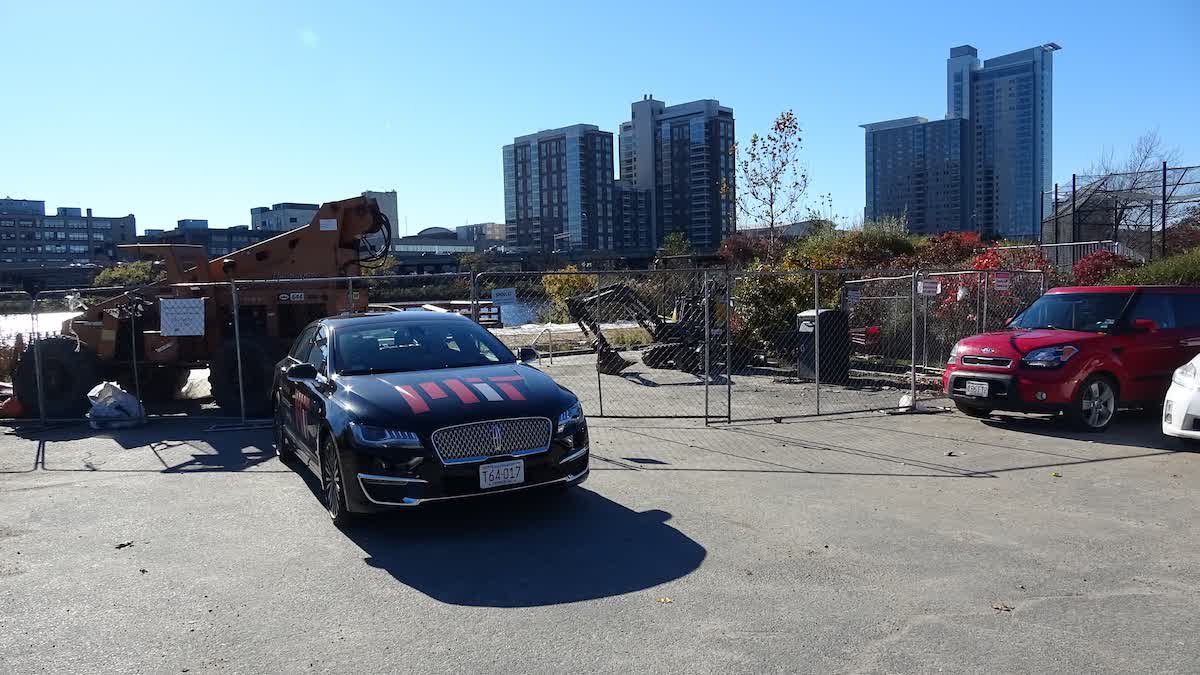}
  
  \lightvspace
  \lightsub{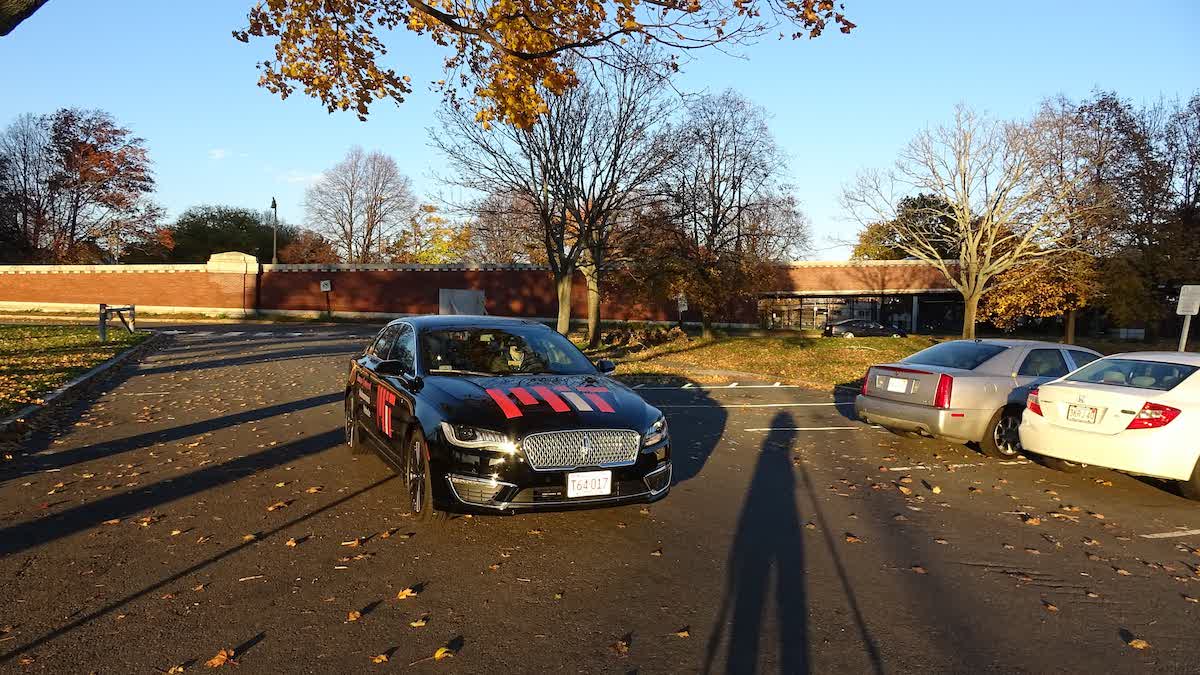}
  \lightsub{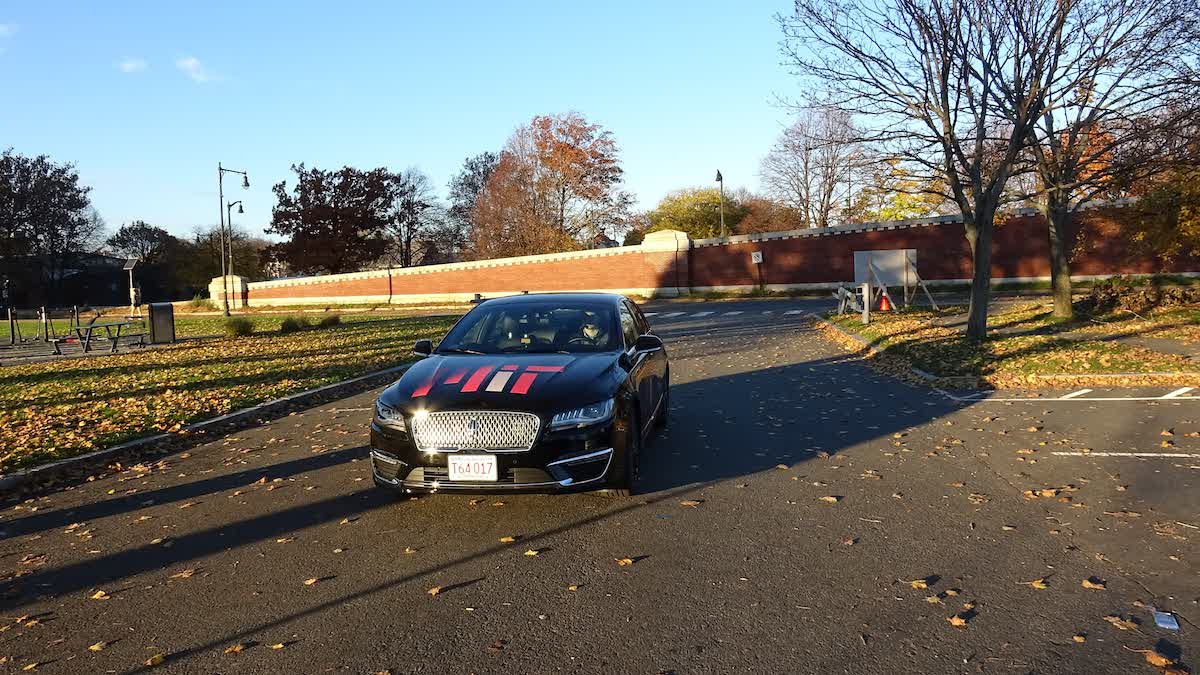}
  \lightsub{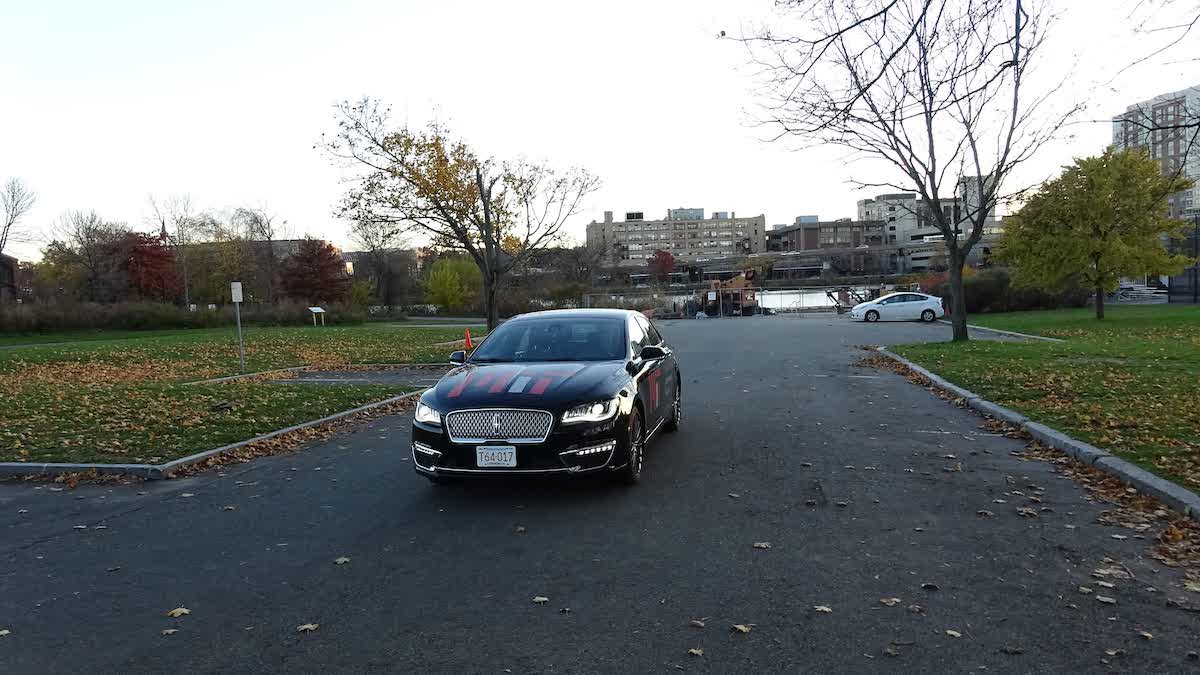}

  \lightvspace
  \lightsub{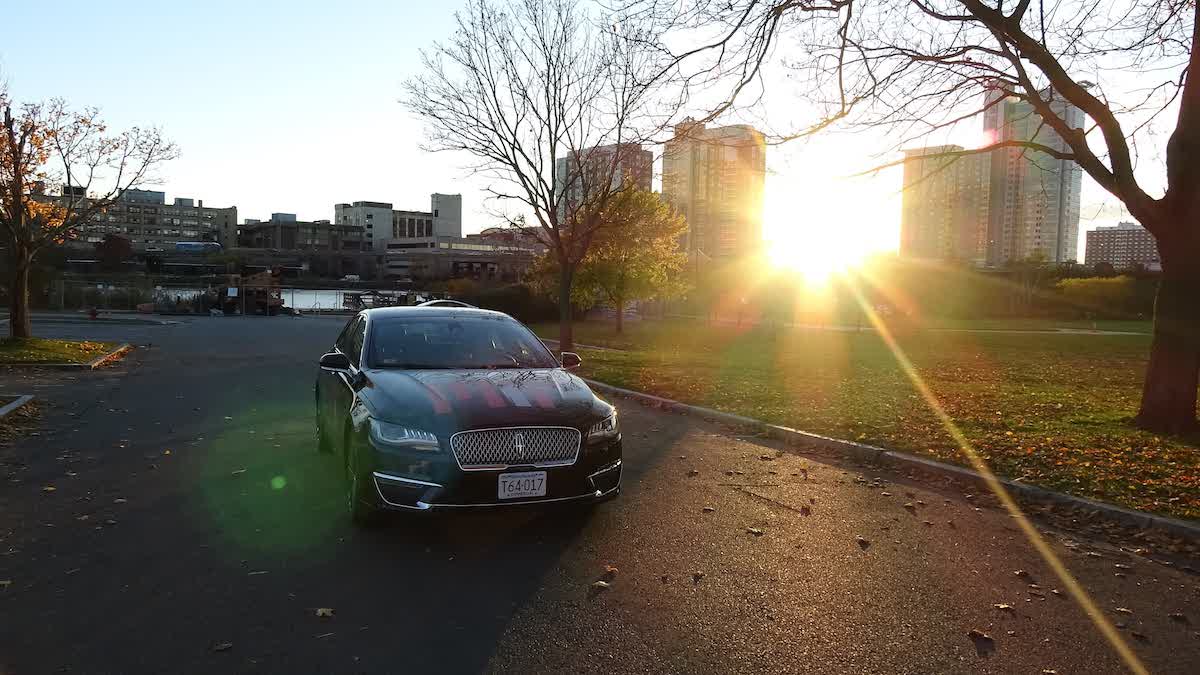}
  \lightsub{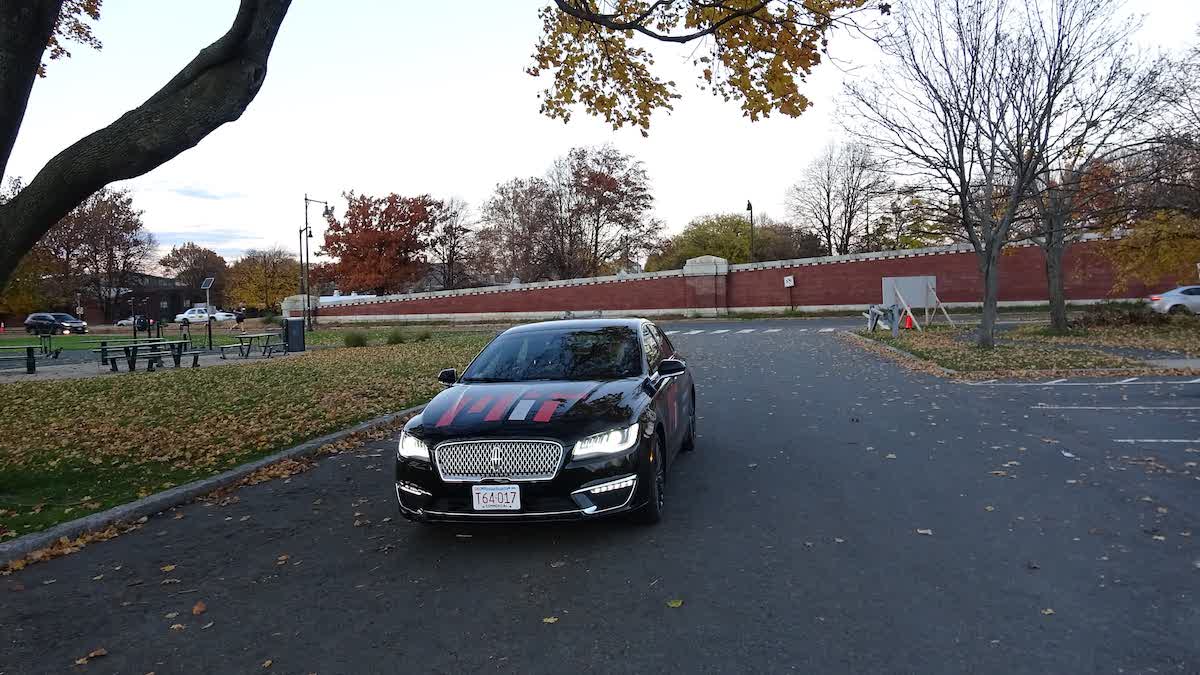}
  \lightsub{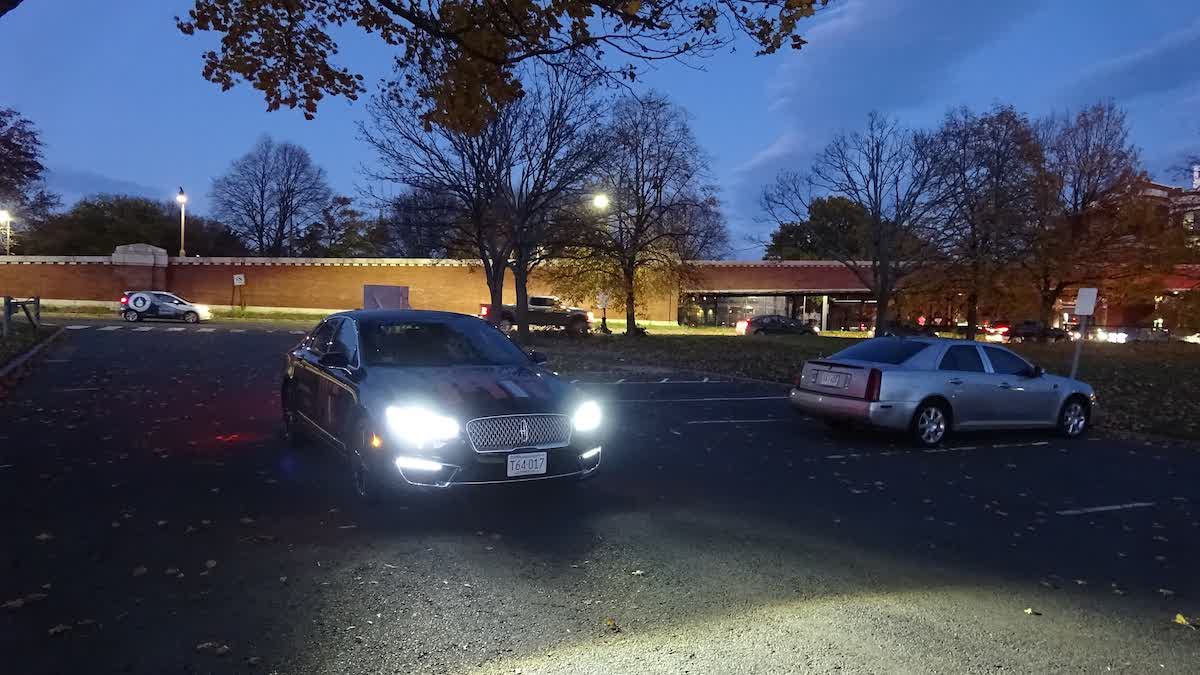}
\caption{Sample stimuli images of approaching vehicle in 9 different lighting conditions.}
\label{fig:lighting}
\end{figure*}

\section{Related Work}\label{sec:relatedwork}

Eye contact is considered to be one of the main non-verbal cues exchanged in driver-pedestrian crossing
interactions. For pedestrians, making eye contact with the driver may facilitate the assessment of drivers' awareness of
their presence or serve as a signal to communicate their intent to cross \citep{rasouli2017agreeing}. With the
introduction of full autonomous cars and the elimination of driver-provided cues, a lot of effort is being directed
towards the understanding of the role of non-verbal cues such as driver presence and eye contact.

\subsection{Driver Presence}

Testing the impact of driver absence on pedestrians' crossing decisions is so far limited to Wizard-of-Oz studies due to
the limited possibility of operating driverless vehicles in urban environments. In this study
\citep{rothenbucher2016ghost}, most pedestrians managed to make crossing decisions in front of a seemingly driverless
car based on vehicle cues alone. 80\% of the interviewed pedestrians noticed the missing driver. It's however unclear
when they noticed it and how it affected their crossing decision as the interviews were conducted after the
interaction. \citep{lundgren2017will} installed a dummy wheel in a right-hand steered vehicle, where the real steering
wheel was hidden from pedestrians. In contrast to the study in \citep{rothenbucher2016ghost}, pedestrians highlighted the
necessity of driver-centric cues, as many indicated they would not cross in front of a driverless vehicle due to lack of
confirmation that they were seen. Conversely, most participants were willing to cross when the driver engaged in eye
contact.

\subsection{Eye Contact}

Eye contact has long been seen as a central component of non-verbal communication in the context of pedestrian-vehicle
interaction. For example, the U.S. Department of Transportation recommends pedestrians to seek eye-contact with drivers
to confirm, that they are seen. However, lost in the term ``eye contact'' is the distinction between (1) looking to
communicate an intent and (2) looking to see. The two goals are disjoint, and the degree to which each is involved in
pedestrian-vehicle communication is important to understand and what our work aims to provide insights on.

\citep{rasouli2017agreeing}, for instance, show that gaze is a powerful tool pedestrians use to communicate their
crossing intent to drivers. They also suggest that pedestrians often establish eye contact with drivers to make sure
they are seen. \citep{sucha2017pedestrian} argue that drivers usually yield if there are pedestrians present at the curb
who clearly communicate their intention of crossing often by seeking eye contact. \citep{gueguen2015pedestrian} also
report that significantly more drivers stopped at a crosswalk when a pedestrian looked them in the eyes in comparison to
when they looked above their head.

In contrast to the common idea about the relevance of eye contact in pedestrian-vehicle interactions,
\citep{dey2017pedestrian} argue that cars' body language - encoded in the kinematics of an approaching vehicle - is much
more significant to traffic negotiations between pedestrians and drivers. They suggest, that pedestrians only seek
eye-contact as a confirmation of driver awareness, if the car ``behaves'' different from what they expected.



To the best of our knowledge, what is not shown in any of the above studies is the distinction between a pedestrian's
behavior of ``seeking eye contact'' and actually being able to see the driver or the driver's eyes. We believe that the latter
is not feasible in most scenarios of an approaching car, but instead, the act of seeking eye contact in itself is used
as social signaling \citep{gobel2015dual}. This distinction is one our work seeks to highlight as we believe it is of
critical importance to the design of autonomous vehicles that safely and effectively interact with pedestrians.

\begin{figure*}[!ht]
  \captionsetup[subfigure]{labelformat=empty}
  \begin{subfigure}{0.49\textwidth}
    \begin{subfigure}{0.48\textwidth}
      \includegraphics[width=\linewidth]{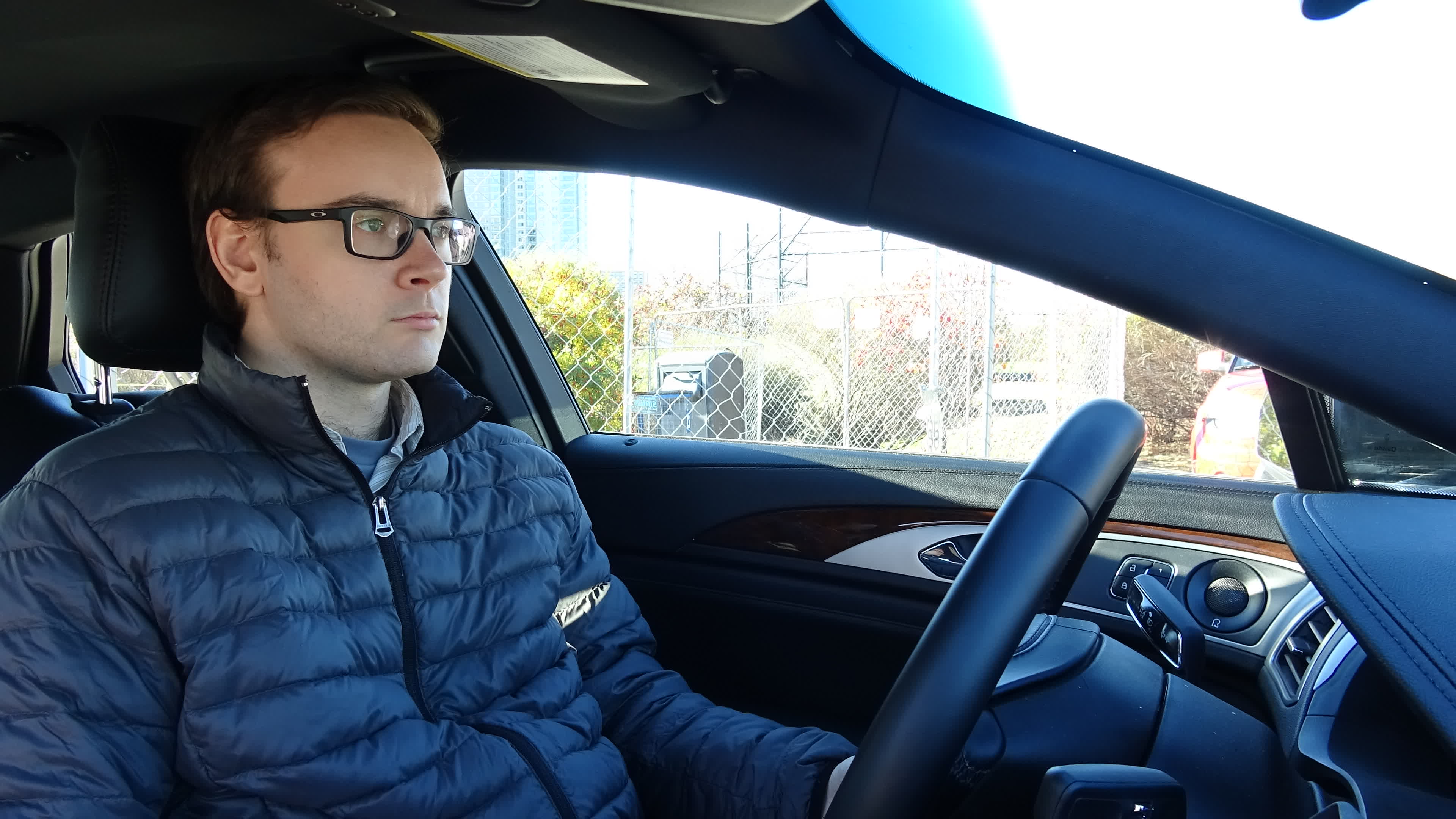}
    \end{subfigure}
    \begin{subfigure}{0.48\textwidth}
      \includegraphics[width=\linewidth]{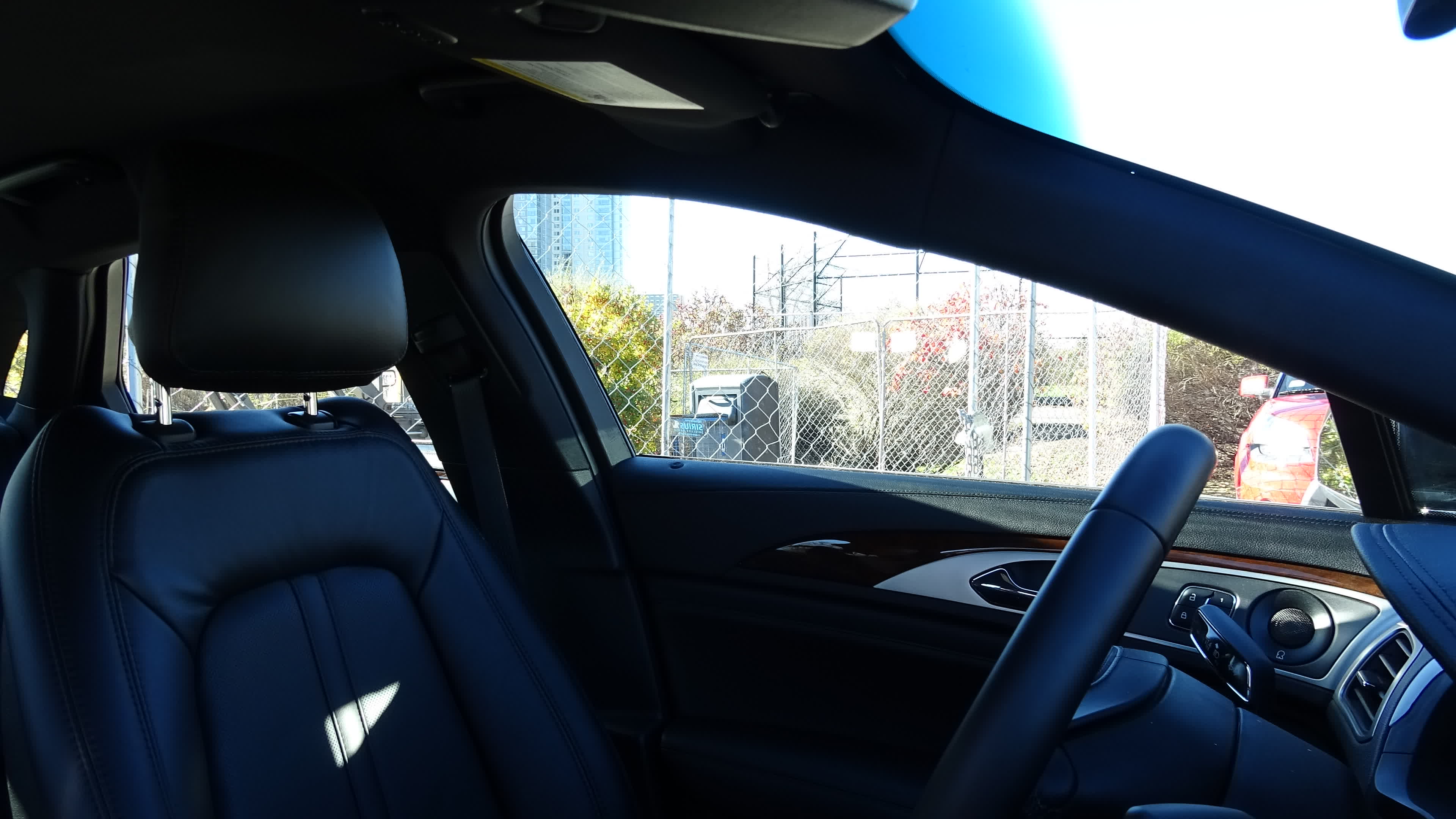}
    \end{subfigure}
    \caption{\textbf{Experiment 1}:\\\ Driver vs. No Driver\\\ \textit{Can you see if there is a driver?}}
  \end{subfigure}
  \begin{subfigure}{0.49\textwidth}
    \begin{subfigure}{0.48\textwidth}
      \includegraphics[width=\linewidth]{images/position_1.jpg}
    \end{subfigure}
    \begin{subfigure}{0.48\textwidth}
      \includegraphics[width=\linewidth]{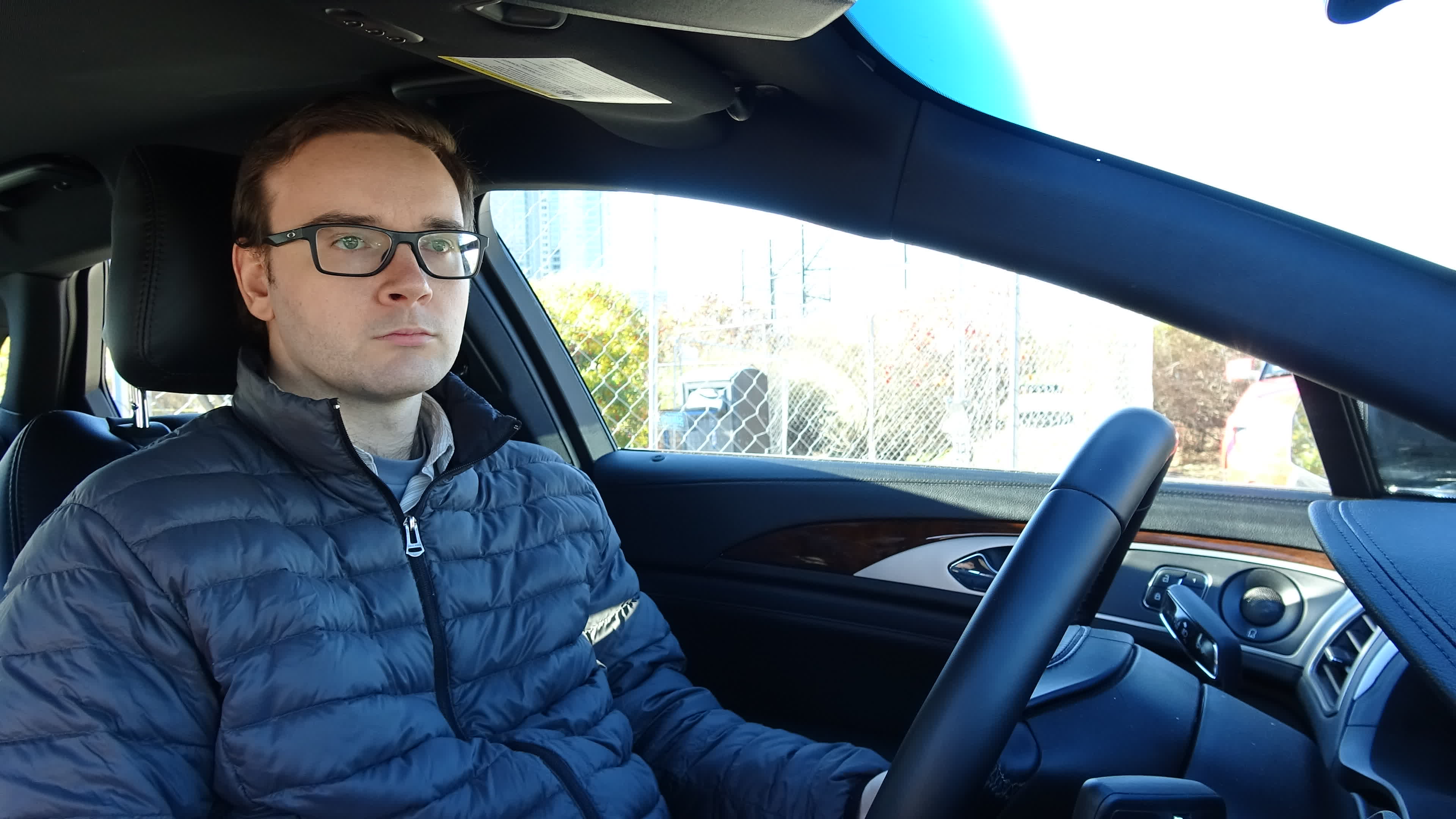} 
    \end{subfigure}
    \caption{\textbf{Experiment 2}:\\\ Looking forward vs. Looking to pedestrian\\\ \textit{Can you see where the driver's eyes are looking?}}
  \end{subfigure}
  \caption{An in-cab view of the driver's body and head position used for the two experiments.}\label{fig:driverstate}
\end{figure*}

\section{Methods}\label{sec:methods}

We conducted 2 experiments to investigate humans' ability to see inside cars. Both experiments were programmed using
\texttt{jsPsych},  a JavaScript library for running behavioral experiments in a web browser and run on Amazon's crowdsourcing
platform Mechanical Turk.

In the first experiment (DRIVER) we examined whether people are able to perceive driver's presence in the car under different
lighting conditions and at different distances. The second experiment (EYES) aimed at identifying the influence of
lighting conditions and vehicle's distance on pedestrians' perception of driver eye-gaze.

\subsection{Stimuli}

The stimuli in both experiments were 4K photos (3840$\times$2160) of a Lincoln MKZ car, that were taken using a Sony
FDR-AX53. We used a tripod at adult eye level (about 165cm) and took the images from the left or the right side of the
vehicle to reflect pedestrians' viewing angle on oncoming cars in a crossing interaction.

The stimuli dataset includes images of the vehicle in \textbf{9 lighting conditions}:
\begin{enumerate*}
\item sunshine (left-view),
\item sunshine (right-view),
\item sun at zenith,
\item sunset (left-view),
\item sunset (right-view),
\item shadow,
\item glare,
\item tree shadow and
\item night
\end{enumerate*} 
(see \figref{lighting}). For each lighting condition, images are taken at \textbf{6 distances}: 5m, 10m, 15m, 20m, 25m and
30m. And for each distance and lighting condition, images are taken featuring \textbf{3 driver states}: looking forward,
looking to pedestrian, absent (see \figref{driverstate}).

\subsection{Experiments}

Both experiments followed a mixed factorial design with two independent variables: \textit{Distance} as the
within-subjects factor and \textit{lighting} as the between-groups factor. Participants were assigned to a random
lighting condition and performed a repeated measurement task for all distances in that particular lighting condition.

\subsubsection{Task}

At the beginning of an experiment, we collected data on gender and age and asked 2 survey questions:

\begin{itemize}
\item \textbf{Q1:} When crossing the street as a pedestrian, when there is a crosswalk and there are no traffic lights,
  do you try to make eye contact with the driver?
\item \textbf{Q2:} Do you believe pedestrians are able to see through a car's windshield and make eye contact with the driver?
\end{itemize}

Then we showed a sample image of the stimuli dataset at a 5m distance and instructed participants to only focus on that
specific car and ignore any other parked cars in the scene. The stimuli of the respective experiment were then presented to the
participant in a random order accompanied by a question for each stimulus. Stimuli for one group were all in one lighting condition and 1 out of 6 distances. For each distance, 2 instances were shown of the driver states of interest to the particular experiment. After the 12 trials, we
asked the participant Q2 again.

\subsubsection{Experiment 1: DRIVER}

In this experiment we included 6 distances and 2 stimuli for each distance: one where the driver is looking forward and one where the
driver is absent. For each image, participants were asked to answer the question:

\textit{Can you see, if there is a driver in the car?}
\begin{itemize}
\item {[n]} No, I can't see if there is a driver.
\item {[0]} Yes, I can see. There is \textbf{no} driver.
\item {[1]} Yes, I can see. There \textbf{is} a driver.
\end{itemize}
 
\subsubsection{Experiment 2: EYE}

In this experiment we included 6 distances and 2 stimuli for each distance: one where the driver is looking forward and one where the
driver is looking to the camera. For each image, participants were asked to answer the question:

\textit{Can you see where the driver's eyes are looking?}
\begin{itemize}
\item {[n]} No, I can't see where the driver's eyes are looking.
\item {[f]} Yes, I can see. They are looking forward.
\item {[c]} Yes, I can see. They are looking to the camera.
\end{itemize}

\subsection{Data collection}

For each experiment subjects were recruited voluntarily through Mechanical Turk. Each experiment involved 180
participants (DRIVER: 79 females and 101 males, EYES: 83 females and 97 males). Their ages ranged from 19 to 74 (M = 39,
SD = 12) for the DRIVER experiment and from 21 to 66 (M = 39, SD = 11) for the EYES experiment. Participants for each
experiment were randomly assigned to one out of 9 lighting condition treatments. Hence, we had 20 participants for each
lighting condition in each experiment. Subjects were rewarded 1.50 USD for participating in the EYES experiment and 0.75
USD for the DRIVER experiment. We only accepted workers who had a minimum of 1000 accepted HITs and a minimum of 98\%
acceptance rate. Furthermore, we measured the response times to the individual stimuli as well as the duration of the
experiment and excluded data from 3 participants: 1 in the DRIVER experiment and 2 in the EYE experiment based on their
very short response times (Mean response time $<$ 400ms).

\newcommand{\commwidth}{0.46}
\begin{figure*}[!ht]
  \centering
  \begin{subfigure}{\commwidth\textwidth}
    \includegraphics[width=\linewidth]{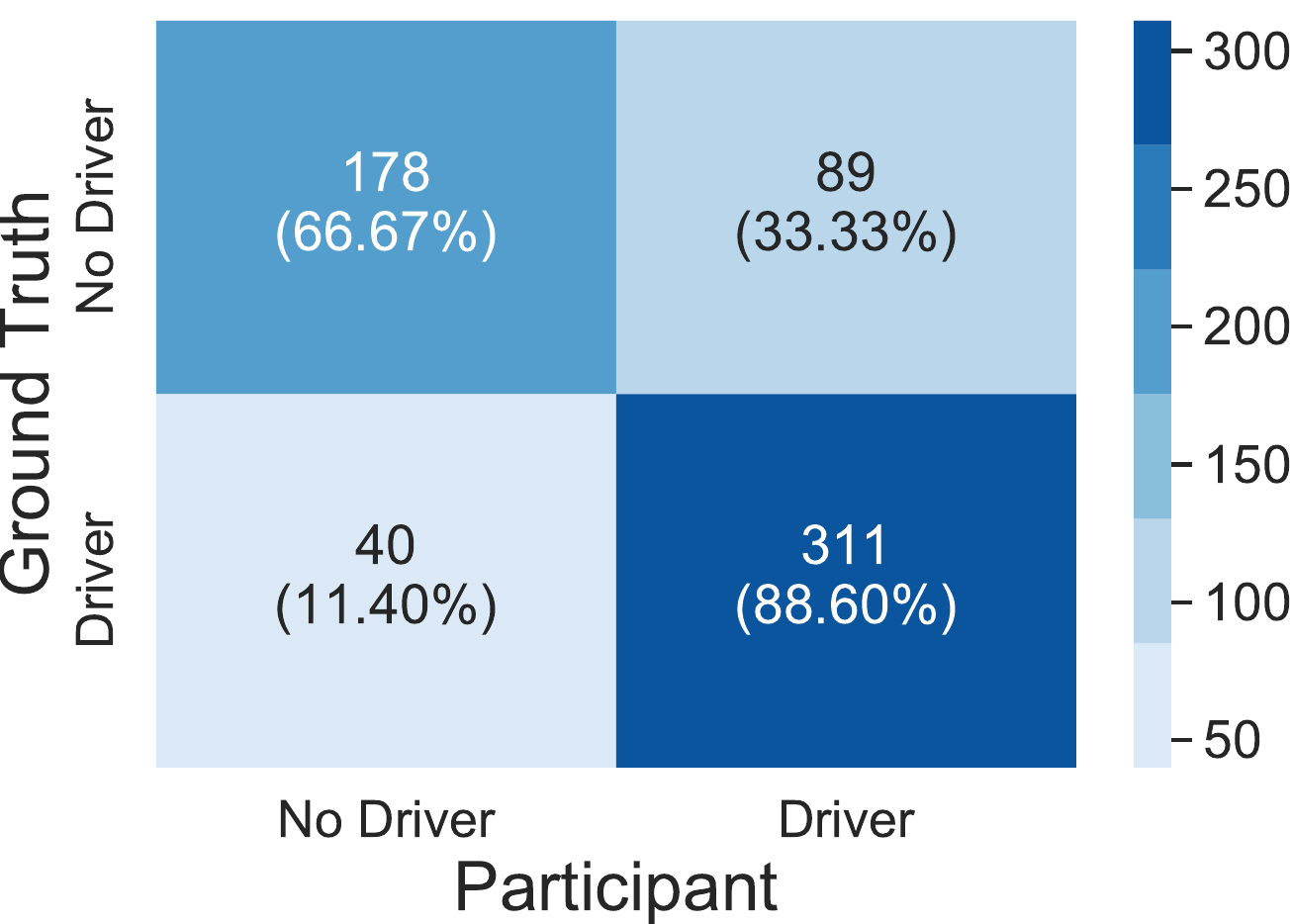}
    \caption{Experiment 1: DRIVER}
  \end{subfigure}
  \hfill
  \begin{subfigure}{\commwidth\textwidth}
    \includegraphics[width=\linewidth]{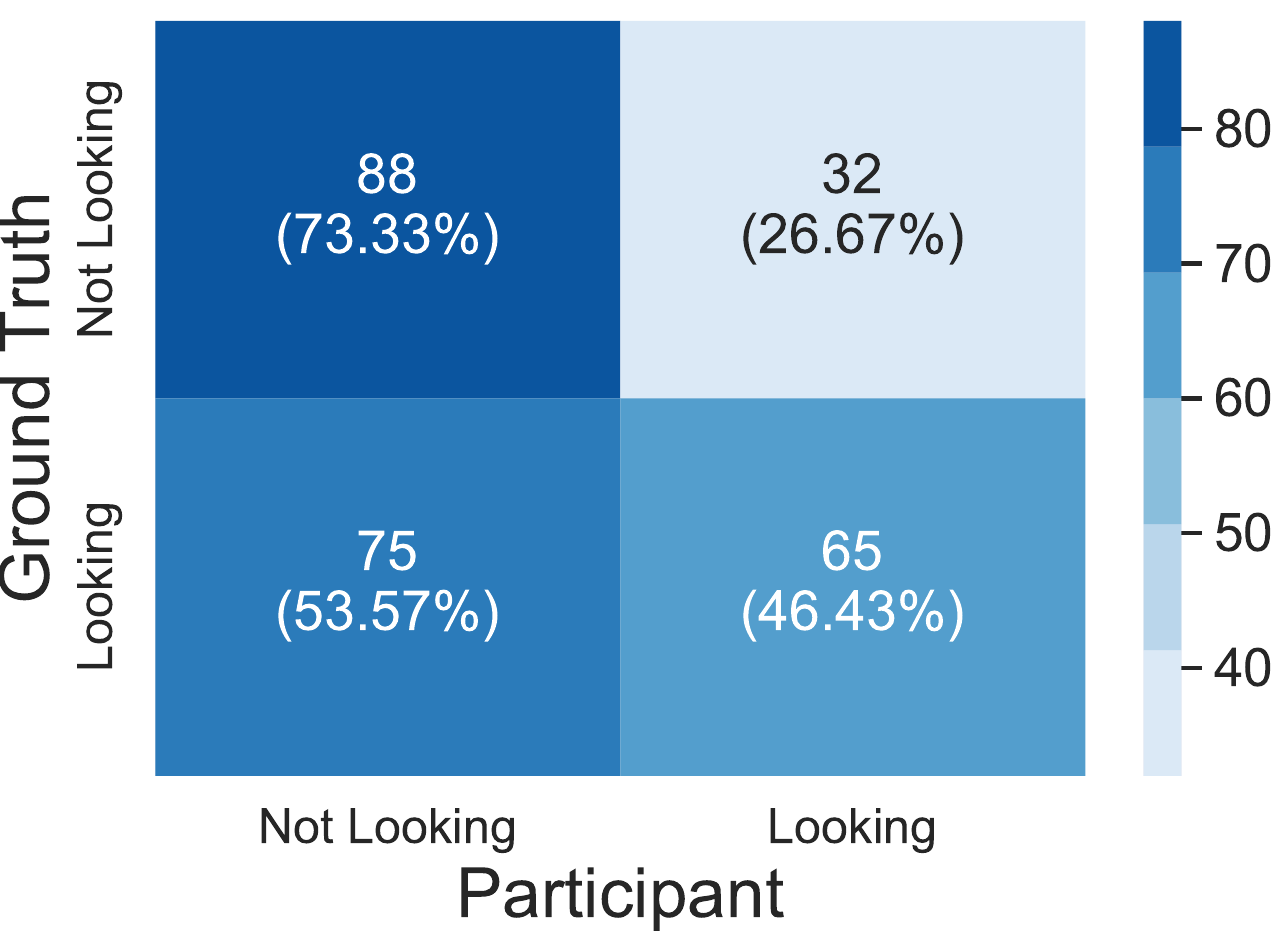}
    \caption{Experiment 2: EYES}
  \end{subfigure}
  \caption{Predictions of driver's presence and eye-gaze in the subset of cases where subject claimed to be able to see
    well enough to answer the question accurately, and then doing so erroneously a large percent of the time, especially
    for the eye-gaze detection experiment.}
  \label{fig:confusion}
\end{figure*}

\begin{figure*}[!ht]
  \vspace{0.2in}
  \begin{subfigure}{\commwidth\textwidth}
    \includegraphics[width=\linewidth]{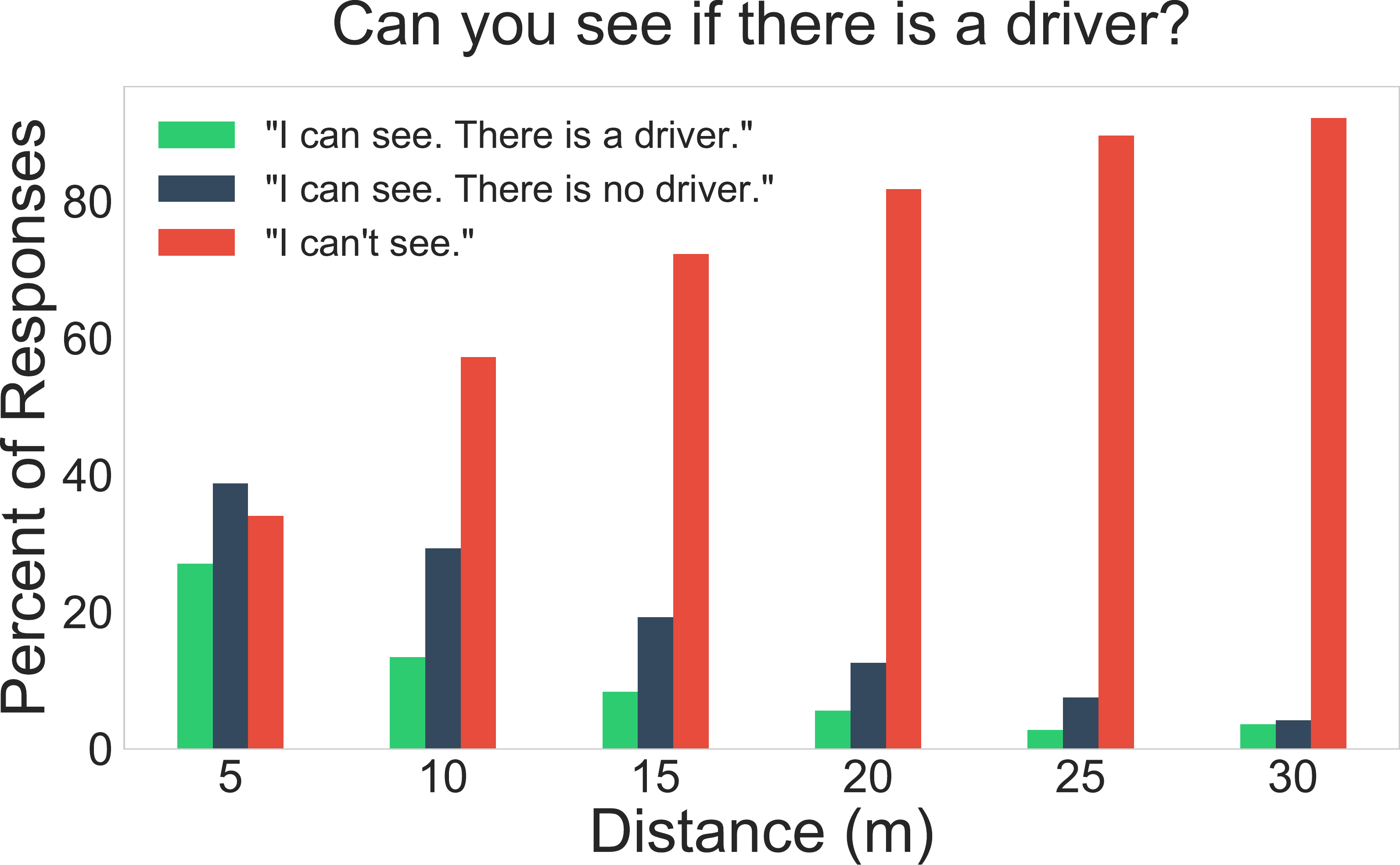}
    \caption{DRIVER: grouped by distance}
  \end{subfigure}\hfill
  \begin{subfigure}{\commwidth\textwidth}
    \includegraphics[width=\linewidth]{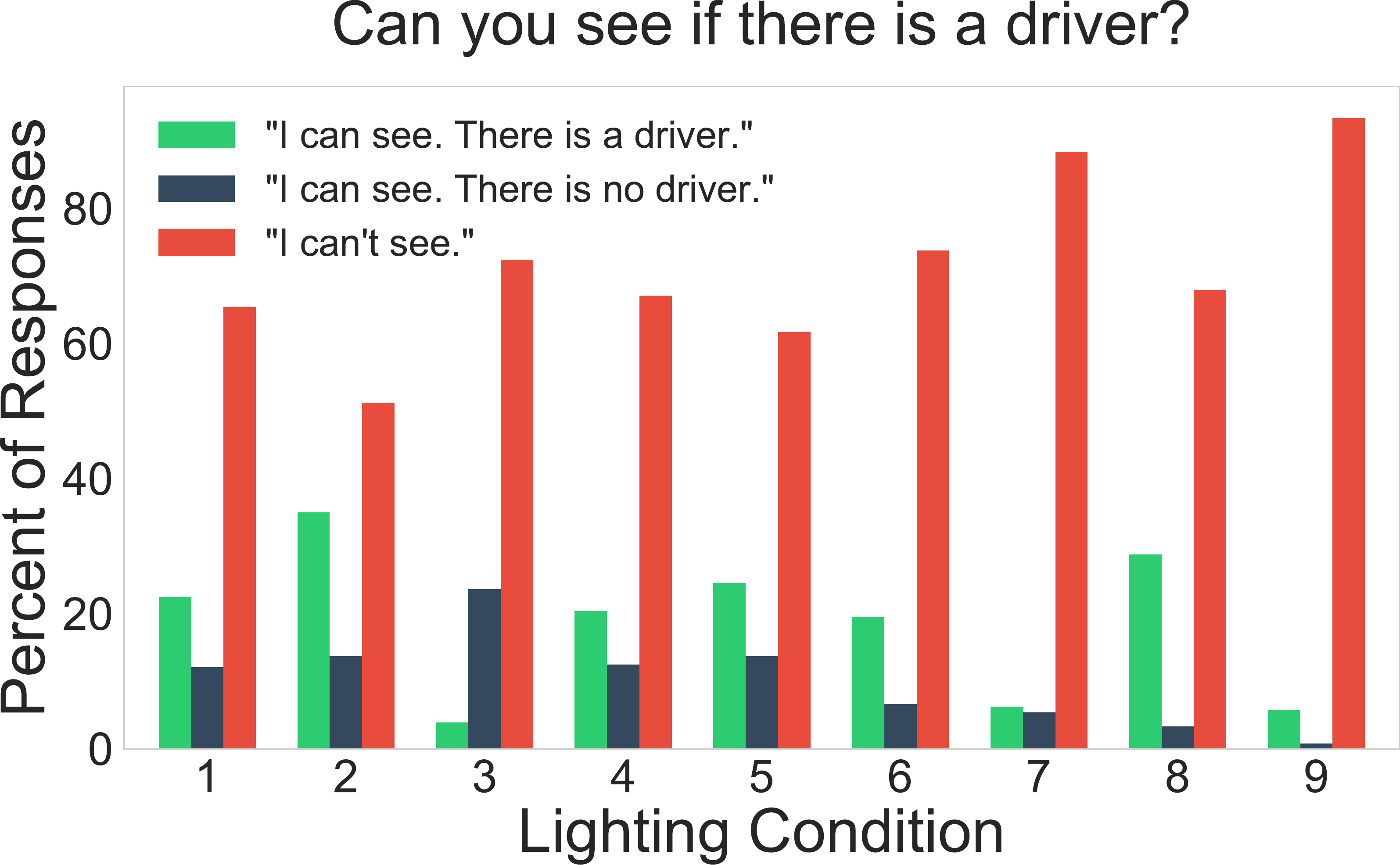}
    \caption{DRIVER: grouped by lighting}
  \end{subfigure}
  \caption{Effect of distance and lighting on perception of driver's presence.}
  \label{fig:driver}
\end{figure*}

\begin{figure*}[!ht]
  \vspace{0.2in}
  \begin{subfigure}{\commwidth\textwidth}
    \includegraphics[width=\linewidth]{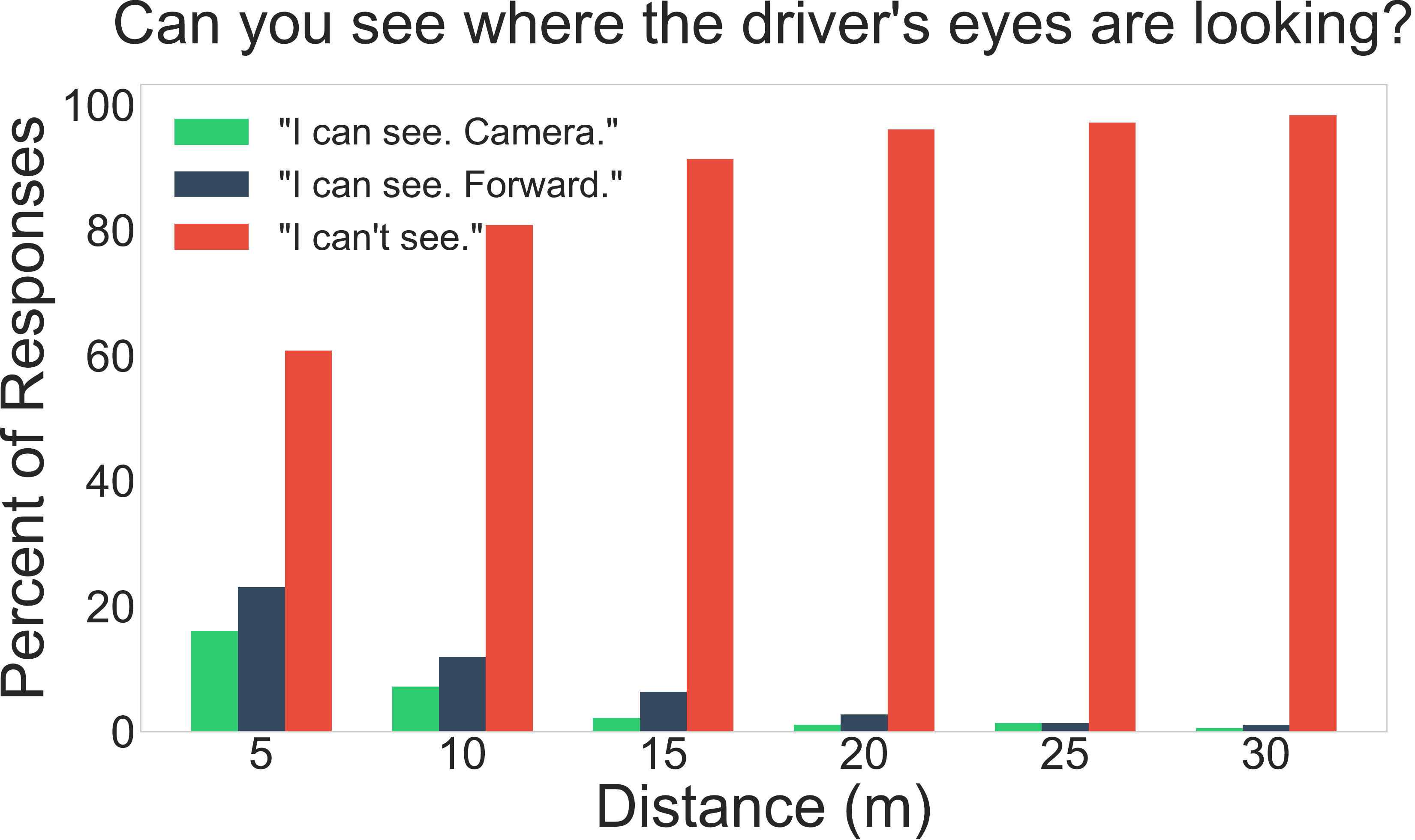}
    \caption{EYES: grouped by distance}
  \end{subfigure}\hfill
  \begin{subfigure}{\commwidth\textwidth}
    \includegraphics[width=\linewidth]{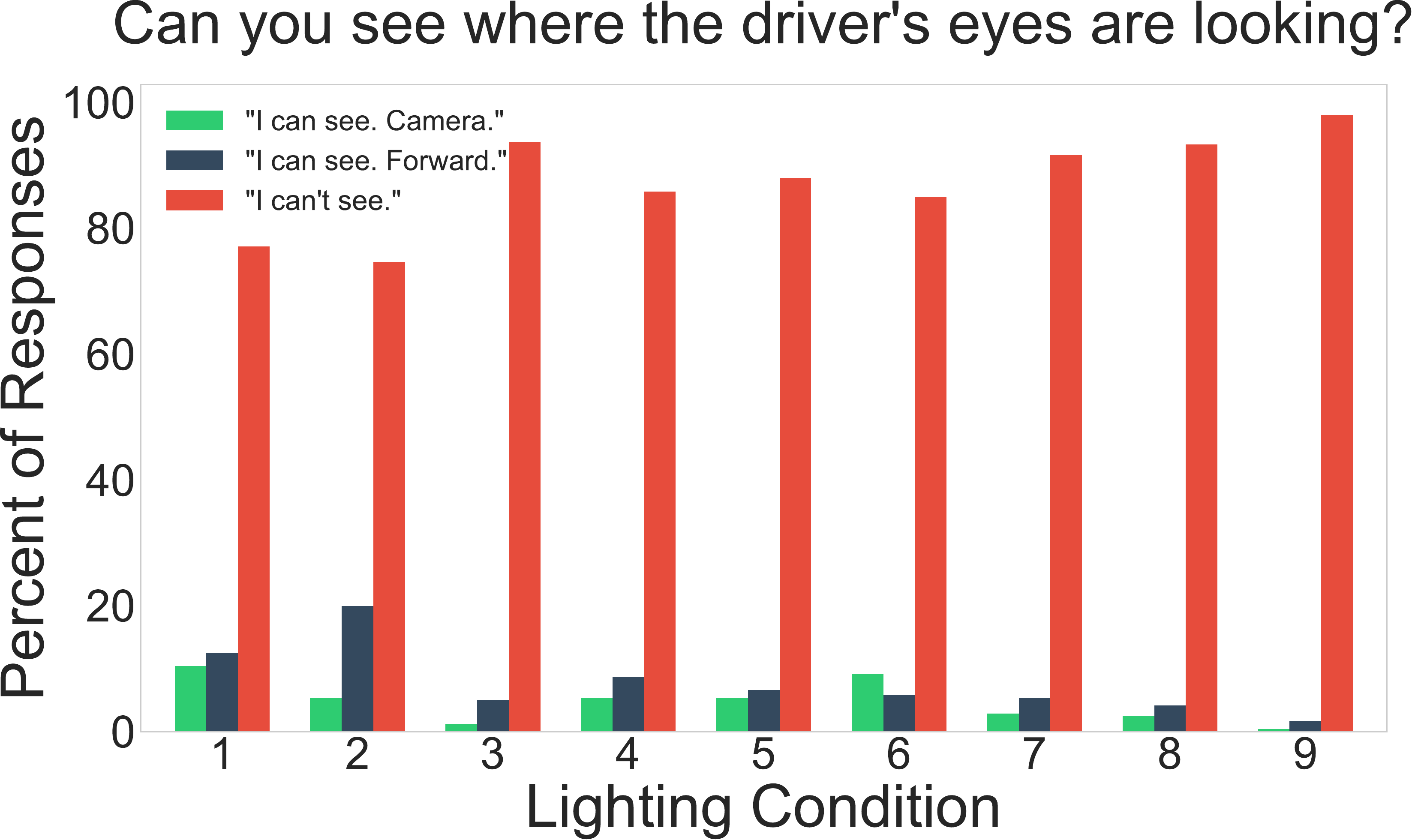}
    \caption{EYES: grouped by lighting}
  \end{subfigure}
  \caption{Effect of distance and lighting on perception of driver's eye-gaze.}
  \label{fig:eye}
\end{figure*}

\section{Results}\label{sec:results}

\subsubsection{Survey}\label{sec:survey}

Prior to the experiment, when asked whether the subject usually seeks eye contact with the driver when crossing the
street (Q1), 122 (34\%) said they don't seek eye contact while 235 (66\%) indicated they do. This confirms our
assumption about the relevance of eye contact as a non-verbal communication mechanism in traffic. To the question about
the feasibility of such communication through the car's windshield (Q2), 282 (79\%) subjects said they believe they can
see through cars' windshields and engage in making eye contact with the driver whereas only 75 (21\%) believed they are
not able to do that. Interestingly, when asked the same question (Q2) again after the experiment, we noticed a decrease
in the total number of people saying they believe they are able to see through windshields from 282 (79\%) to 190
(53\%). Particularly, one third of the participants (104) change their response from ``Yes, I am able to see through
windshields'' to ``No, I am not able to see through windshields'' after the experiment.

\subsubsection{Experiment}

\begin{table}[!ht]
  \begin{center}
    \begin{tabular}{ |c||c|c|  }
      \hline
      &Number of responses & Number of ``I can't see''\\
      \hline
      DRIVER   & 2148   &1530 (71\%)\\
      EYES&   2136  & 1876 (87\%)\\
      \hline
    \end{tabular}
  \end{center}
  \caption{Experiment results}
  \label{table}
\end{table}

Table~\ref{table} shows that 71\% of the responses to stimuli were \textit{I can't see}, meaning that in all those
instances subjects were unable to see whether there was a driver or not. In the EYES experiment, subjects were not able
to detect the driver's eye-gaze in 87\% of the instances. Decisive responses, where subjects made a prediction, had an
accuracy of 0.8 for the DRIVER experiment and 0.6 for the EYES experiment (see \figref{confusion}).

In \figref{driver} and \figref{eye} we group the responses by the distances and lighting conditions. For all lighting
conditions and all distances except at 5m distance, the highest number of responses to the stimuli is \textit{I can't
  see}. The percentage of such uncertain responses is around 60\% at 10m and increases with more distance to reach 90\%
of all responses at a 25m distance in the DRIVER experiment. For the perception of driver's eye-gaze, responses indicate
that participants can't see whether a driver is looking at them at a 5m distance (60\% I can't see). This rises up to
91\% of responses at 15m distance. These results show that seeing through cars' windshields is very limited and can only
occur at very short distances to the vehicle. When applied to the crossing context, this implies that crossing
pedestrians don't mainly rely on driver-provided cues (eye contact) as those usually occur when the decision to cross
has already been made. For example, assuming a speed of 20mph, the time to collision at 30m is 3.3s which is below
pedestrians' threshold for crossing. Hence, a pedestrian must have made their decision to cross before the car reaches
30m distance at that speed. Pedestrians are therefore not necessarily basing their decision of crossing on the
perception of driver-related cues.

Based on this data, we conclude that pedestrians are often unable to see through the windshield and engage in making eye
contact with the driver. According to the answers to Q1 and Q2, however, a common contradicting belief exists about our
ability to make eye contact. Assuming pedestrians look towards the windshield when communicating with the driver,
perhaps there is something functionally similar to an eye-gaze that provides them with acknowledgment and therefore
simulates engaging in eye contact with the driver.

The discrepancy between the reports of pedestrians and these results may be that people create post-hoc explanations of
their decisions \citep{Nisbett1977}. The perceived coupling of the kinematic behavior with pedestrians' gaze behavior
may lead them to believe that their gaze resulted in the vehicle stopping.  Overall, the results speak to the larger
challenge of relying on reports from people in the design of automated vehicles for tasks that involve tacit knowledge
or high degrees of automaticity \citep{Nisbett1977}. Even subtle influences such as question wording have been shown to
change how people report their experiences \citep{Loftus1975}. These results in the larger context emphasize the
importance of data when making decisions about the design of safety-relevant systems.

\section{Conclusion}\label{sec:conclusion}

In this paper we examine the role of eye contact and driver presence in vehicle-pedestrian interactions. We challenge
the common notion that decision making in crossing interactions is largely influenced by direct human-to-human
communication between pedestrians and drivers. We show that over 90\% of people cannot determine the gaze of the driver
at 15m and see the driver at all at 30m. This means that, for example, given the common city speed limit of 25mph, more
than 99\% of pedestrians would have begun crossing an unsignalized crosswalk before being able to see either the driver or the driver's gaze
\citep{brewer2006exploration}. In other words, from the perspective of the pedestrian, in most situations involving an
approaching vehicle, the crossing decision is made by the pedestrian solely based on the kinematics of the vehicle
without needing to determine that eye contact was made by explicitly detecting the eyes of the driver.

\section*{Acknowledgment}

This work was in part supported by US DOT's Region I New England University Transportation Center at MIT (NEUTC) and
Toyota's Collaborative Safety Research Center. The views and conclusions being expressed are those of the authors and
may not necessarily reflect the views of sponsoring organizations.


\balance

\bibliography{inside}

\begin{thebibliography}{11}
\providecommand{\natexlab}[1]{#1}
\providecommand{\url}[1]{\texttt{#1}}
\expandafter\ifx\csname urlstyle\endcsname\relax
  \providecommand{\doi}[1]{doi: #1}\else
  \providecommand{\doi}{doi: \begingroup \urlstyle{rm}\Url}\fi

\bibitem[Brewer et~al.(2006)Brewer, Fitzpatrick, Whitacre, and
  Lord]{brewer2006exploration}
M.~A. Brewer, K.~Fitzpatrick, J.~A. Whitacre, and D.~Lord.
\newblock Exploration of pedestrian gap-acceptance behavior at selected
  locations.
\newblock \emph{Transportation research record}, 1982\penalty0 (1):\penalty0
  132--140, 2006.

\bibitem[Dey and Terken(2017)]{dey2017pedestrian}
D.~Dey and J.~Terken.
\newblock Pedestrian interaction with vehicles: roles of explicit and implicit
  communication.
\newblock In \emph{Proceedings of the 9th International Conference on
  Automotive User Interfaces and Interactive Vehicular Applications}, pages
  109--113. ACM, 2017.

\bibitem[Fridman(2018)]{fridman2018humancentered}
L.~Fridman.
\newblock Human-centered autonomous vehicle systems: Principles of effective
  shared autonomy.
\newblock \emph{CoRR}, abs/1810.01835, 2018.
\newblock URL \url{https://arxiv.org/abs/1810.01835}.

\bibitem[Gobel et~al.(2015)Gobel, Kim, and Richardson]{gobel2015dual}
M.~S. Gobel, H.~S. Kim, and D.~C. Richardson.
\newblock The dual function of social gaze.
\newblock \emph{Cognition}, 136:\penalty0 359--364, 2015.

\bibitem[Gu{\'e}guen et~al.(2015)Gu{\'e}guen, Meineri, and
  Eyssartier]{gueguen2015pedestrian}
N.~Gu{\'e}guen, S.~Meineri, and C.~Eyssartier.
\newblock A pedestrian’s stare and drivers’ stopping behavior: A field
  experiment at the pedestrian crossing.
\newblock \emph{Safety science}, 75:\penalty0 87--89, 2015.

\bibitem[Loftus and Zanni(1975)]{Loftus1975}
E.~F. Loftus and G.~Zanni.
\newblock {Eyewitness testimony: The influence of the wording of a question}.
\newblock \emph{Bulletin of the Psychonomic Society}, 5\penalty0 (1):\penalty0
  86--88, jan 1975.
\newblock \doi{10.3758/BF03336715}.

\bibitem[Lundgren et~al.(2017)Lundgren, Habibovic, Andersson, Lagstr{\"o}m,
  Nilsson, Sirkka, Fagerl{\"o}nn, Fredriksson, Edgren, Krupenia,
  et~al.]{lundgren2017will}
V.~M. Lundgren, A.~Habibovic, J.~Andersson, T.~Lagstr{\"o}m, M.~Nilsson,
  A.~Sirkka, J.~Fagerl{\"o}nn, R.~Fredriksson, C.~Edgren, S.~Krupenia, et~al.
\newblock Will there be new communication needs when introducing automated
  vehicles to the urban context?
\newblock In \emph{Advances in Human Aspects of Transportation}, pages
  485--497. Springer, 2017.

\bibitem[Nisbett and Wilson(1977)]{Nisbett1977}
R.~E. Nisbett and T.~D. Wilson.
\newblock {Telling more than we can know: Verbal reports on mental processes.}
\newblock \emph{Psychological Review}, 84\penalty0 (3):\penalty0 231--259,
  1977.
\newblock \doi{10.1037/0033-295X.84.3.231}.

\bibitem[Rasouli et~al.(2017)Rasouli, Kotseruba, and
  Tsotsos]{rasouli2017agreeing}
A.~Rasouli, I.~Kotseruba, and J.~K. Tsotsos.
\newblock Agreeing to cross: How drivers and pedestrians communicate.
\newblock In \emph{Intelligent Vehicles Symposium (IV), 2017 IEEE}, pages
  264--269. IEEE, 2017.

\bibitem[Rothenb{\"u}cher et~al.(2016)Rothenb{\"u}cher, Li, Sirkin, Mok, and
  Ju]{rothenbucher2016ghost}
D.~Rothenb{\"u}cher, J.~Li, D.~Sirkin, B.~Mok, and W.~Ju.
\newblock Ghost driver: A field study investigating the interaction between
  pedestrians and driverless vehicles.
\newblock In \emph{Robot and Human Interactive Communication (RO-MAN), 2016
  25th IEEE International Symposium on}, pages 795--802. IEEE, 2016.

\bibitem[Sucha et~al.(2017)Sucha, Dostal, and Risser]{sucha2017pedestrian}
M.~Sucha, D.~Dostal, and R.~Risser.
\newblock Pedestrian-driver communication and decision strategies at marked
  crossings.
\newblock \emph{Accident Analysis \& Prevention}, 102:\penalty0 41--50, 2017.

\end{thebibliography}

\end{document}